\pdfoutput=1
\documentclass[5p,times]{elsarticle}
\journal{Thin Solid Films}

\usepackage[T1]{fontenc}
\usepackage{textcomp}
\usepackage{amsmath}
\usepackage{booktabs}
\usepackage{array}
\usepackage{hyperref}
\hypersetup{colorlinks=true}

\biboptions{sort&compress}

\providecommand*{\rb}[1]{\ensuremath{_\mathrm{#1}}} 
 
\providecommand{\un}[1]{\ensuremath{\,\mathrm{#1}}}

\makeatletter
\providecommand*{\upbox}[1]{\textrm{\upshape#1}}
\providecommand{\C}{\ensuremath{\upbox{\textdegree}\kern-\scriptspace\mathrm{C}}}

\providecommand{\micro}{\upbox{\textmu}}
\providecommand{\ohm}{\upbox{\textohm}}

\newcolumntype{C}{>{$}c<{$}}
\newcolumntype{L}{>{$}l<{$}}
\newcolumntype{R}{>{$}r<{$}}

\makeatother

\begin{document}

\begin{frontmatter}

\title{Impurity effects in Cu$_2$O}

\author[ENEA]{Francesco Biccari\corref{biccari}}
\ead{francesco.biccari@enea.it}

\author[UNITN,ENEA]{Claudia Malerba}

\author[ENEA]{Alberto Mittiga}

\cortext[biccari]{Corresponding author. Tel.: +39\,3048\,6640; fax: +39\,3048\,6405.}
\address[ENEA]{ENEA, Casaccia Research Center, via Anguillarese 301, 00123 Roma, Italy}
\address[UNITN]{University of Trento, Department of Civil, Environmental and Mechanical Engineering, via Mesiano 77, 38123 Trento, Italy}

\begin{abstract}
The doping of wide gap semiconductors is an interesting problem both from the scientific and technological point of view.
A well known example of this problem is the doping of Cu$_2$O.
The only element which has produced an order of magnitude increase in the conductivity of Cu$_2$O bulk samples is chlorine, as previously reported by us and others.
However the solar cells produced with this material do not show any improvement in performances because of the reduction in the minority carrier diffusion length.
In this paper we investigate the effect of other impurities in order to check their possible use as dopants and to assess their effects on the minority carrier diffusion length.
Seven impurities have been introduced by evaporation on the starting copper sheet before the oxidation used to produce Cu$_2$O: chromium (Cr), iron (Fe), silver (Ag), silicon (Si), sodium (Na), sulfur (S) and phosphorus (P).
The experiments show that a 20\,ppm of concentration of these dopants does not give any relevant effect neither on the resistivity, nor on the mobility.
The effect on minority carrier diffusion length is also negligible except for sodium which produces a slight degradation of the samples.
\end{abstract}


\end{frontmatter}

\section{Introduction}

Cuprous oxide (Cu$_2$O) is a spontaneously $p$-type semiconductor with a gap of $2.09\un{eV}$ at room temperature. Both experiments~\cite{Porat_1995} and \textit{ab initio} calculations~\cite{Raebiger_2007, Soon_2009} suggest that the $p$-type conductivity at high temperature is due to the presence of copper vacancies ($V\rb{Cu}$).
Even if the \textit{ab initio} calculations do not predict a donor with a small formation energy, the experiments at low temperature~\cite{Brattain_1951, Mittiga_2006_B} ($T < 450\un{K}$) show that Cu$_2$O is a compensated material with a compensation ratio $N\rb{A}/N\rb{D}$ which varies from values just slightly larger than $1$ to values of the order of $10$~\cite{Brattain_1951,Pollack_1975}.
The nature of the compensating donor is still controversial.
The simplest idea identifies the donor centers with the oxygen vacancies ($V\rb{O}$)~\cite{Bloem_1958}, however a more refined model~\cite{Mittiga_2009} suggests the defect complex ($V\rb{Cu} - V\rb{O}$) as the compensating donor.

Thanks to the value of its energy gap and to the low production cost, cuprous oxide has always been considered a good candidate for low-cost solar cell and recently its use in the top cell of a multi-junction solar cell or as the host material for intermediate band solar cell, was proposed.
Until now no reliable doping processes of bulk Cu$_2$O has been developed and the best solar cells, with an efficiency of about $2\%$, were made using undoped substrates~\cite{Mittiga_2006}.

Since the efficiency of the record solar cell is strongly limited by series resistance effects, the use of substrates with a greater conductivity could improve the device performances.

We want to stress the fact that the doping level required for a solar cell with a suitable resistance of the $p$-type Cu$_2$O is not so difficult to achieve. 
Using an acceptor level located at about $0.2\un{eV}$ above the valence band~\cite{Ishizuka_2003} and
considering the presence of compensating donors in the typical concentration ($10^{14}\un{cm^{-3}}$), it is straightforward to show that an extrinsic acceptor concentration of $10^{16}\un{cm^{-3}}$ can give a negligible voltage drop on the series resistance for a typical Cu$_2$O bulk solar cell.
It is important to note that these concentrations are well below the solubility limits of the impurities in Cu$_2$O~\cite{Tsur_1998,Tsur_1999}.
Therefore the problem is only to find a suitable impurity for Cu$_2$O which introduces a relatively shallow acceptor level without increasing too much the recombination rate for the minority carriers.

Another reason to study the impurity effects is that a large scattering of electrical data exists in the Cu$_2$O literature. This is possibly related to the impurity concentration in the starting copper used for oxidation~\cite{Trivich_1981_tr_final}. Indeed typical intrinsic defect density found in Cu$_2$O is of the order of $10^{15}\un{cm^{-3}}$ while the impurities concentration from the starting copper was always larger than $5 \times 10^{16}\un{cm^{-3}}$ ($1\un{ppm}$).
A systematic study of the effect of the most common impurities in Cu$_2$O is therefore necessary.

\section{Survey of doping in Cu$_2$O}

Looking at the literature on doping of Cu$_2$O thin films, it is often concluded that the doping of Cu$_2$O is a feasible task. On  Cu$_2$O sputtered thin films a reduction of about one order of magnitude of the resistivity was obtained by nitrogen~\cite{Ishizuka_2001}, silicon and germanium~\cite{Ishizuka_2002, Ishizuka_2004} which gave a $p$-type material.
A further conductivity increase, both on doped and undoped material, can be achieved by the passivation of defects by hydrogen and/or cyanide treatments even if a surprising and strong reduction of the mobility is also observed~\cite{Tabuchi_2002, Ishizuka_2002_B, Ishizuka_2003, Okamoto_2003, Lu_2005, Lu_2005_B, Paul_2008, Akimoto_2006}.
Other tested impurities were tin (Sn) and lead (Pb) which behave as donors. However, due to the \emph{self-compensation} effect induced by the copper vacancies formation~\cite{Tsur_1999_B}, they lead only to a more resistive material~\cite{Ishizuka_2004}.
Nickel was also extensively studied in thin films grown by pulsed laser deposition~\cite{Kikuchi_2005, Kikuchi_2006}. It acts as a neutral impurity which can reduce the mobility. This is in contrast with the theoretical calculation~\cite{Martinez-Ruiz_2003}.

As for electrodeposited thin films, the only tested dopant was chlorine~\cite{Han_2009, Han_2010}, which was claimed to give a $n$-type Cu$_2$O.

Finally we have to mention cobalt~\cite{Kale_2003, Antony_2007, Raebiger_2007} and manganese~\cite{Liu_2005, Wei_2005} which were studied in detail because of their magnetic properties. However the doping of Cu$_2$O with these material does not give an appreciable reduction of the resistivity.

The doping of bulk samples is an even more difficult task. No one ever succeeded in obtaining $n$-type doping of a bulk Cu$_2$O. It is generally accepted that $n$-type doping is forbidden in the thermodynamic equilibrium because of a \emph{self-compensation} mechanism~\cite{Tsur_1999_B}.
Even $p$-type doping is not easy.
Some successes were obtained using chlorine~\cite{Olsen_1982, Musa_1998, Biccari_2010} and cadmium~\cite{Tapiero_1979, Papadimitriou_1984_proc, Papadimitriou_1989, Papadimitriou_1989_B, Trivich_1978}.
The former gave a $p$-type material with a minimum resistivity of $66\un{\ohm\,cm}$~\cite{Olsen_1982} while the latter gave, for a very large concentration of cadmium of about 1\%, a minimum resistivity of $9\un{\ohm\,cm}$~\cite{Trivich_1978}. However solar cells made with these doped substrates did not give efficiencies better than those obtained by the undoped material. In the case of chlorine this is due to a reduction of the electron diffusion length.

Several other impurities (Be~\cite{O'Keeffe_1961}, Tl~\cite{Trivich_1978}, Si~\cite{Nemoto_1967}, Ag~\cite{Trivich_1978}, Fe~\cite{Tapiero_1979}, Al and In~\cite{Trivich_1978}, Co~\cite{Tsur_1995}) 
were roughly tested but none of them gave a reduction of resistivity.
At least for aluminium and indium, their ineffectiveness was explained by the formation of neutral defect complexes between these elements and the copper vacancies~\cite{Wright_2002}.

In all these works the Cu$_2$O doping technology was very primitive and the impurity concentration was never accurately controlled.

\section{Preparation of Cu$_2$O doped substrates}

In this paper we investigate the effect of seven different impurities, chromium (Cr), iron (Fe), silver (Ag), silicon (Si), sodium (Na), sulfur (S) and phosphorus (P), from the point of view of the morphology, electrical properties and minority carrier diffusion length.
The choice of these elements was dictated, in addition to their immediate availability in our laboratory, by different reasons.
Chromium and iron were tried to understand if the low resistivity shown by some Trivich's samples~\cite[p.~35]{Trivich_1981_tr_final}, obtained by rolling very pure copper rods between two stainless steel cylinders, is due to the introduction of contaminants from the steel.
He observed that while the rolled copper gave samples with $\rho\approx 100 - 200\un{\ohm\,cm}$, the untreated material gave a Cu$_2$O with a resistivity of the order $10^4\un{\ohm\,cm}$.
Silver and sulfur were tried because they are commonly found in most of the commercial copper.
Moreover silver is the element with the largest solubility in Cu$_2$O~\cite{Tsur_1999} and therefore this trial is the first step of a study of the heavy doping effects in Cu$_2$O which will be described elsewhere.
Sodium was tested because it showed a very positive effect in chalcopyrites, increasing the conductivity and the grain size~\cite{Rau_1999}.
Silicon was tested because it produces $p$-type doping in thin films and finally phosphorus was tested because it is in the fifth group like the nitrogen, which is a good dopant for Cu$_2$O thin films.

In this work we have not tested divalent metals. As discussed in the previous section, some of them can increase the conductivity while others have the opposite effect. According to the simulations this behavior seems to be related to the ionic radius of the impurity~\cite{Nolan_2008}. We plan to test this predictions in a future work.

All dopants have been evaporated on both surfaces of the starting $99.9999\%$ pure copper $100\un{\micro m}$ thick:
for silicon we have evaporated SiO, for the sodium the NaOH, for sulphur the CuS and for phosphorus the P$_2$O$_5$, while the others elements were evaporated from their elemental form.
These dopant layers have been covered by a $100\un{nm}$ thick layer of  evaporated copper in order to avoid the evaporation of the dopant during the oxidation.

The oxidation proceeds as follows:
the temperature is raised with a rate of $10\un{\C/min}$ up to $910\un{\C}$ in nitrogen flux (oxygen impurity content $5\un{ppm}$).
Then a $90\un{min}$ long oxidation in a partial oxygen pressure of $2.7\un{Torr}$ is done.
The oxidation is followed by two annealings: the first one at $1125\un{\C}$ for $60\un{min}$ and the second one at $780\un{\C}$ for $600\un{min}$, both under a $0.27\un{Torr}$ oxygen partial pressure.
Finally, after switching to the nitrogen flux only, the sample is cooled down to $450\un{\C}$ and then quenched to room temperature dipping it into deionized water.

A test with $20\un{ppm}$ of element with respect to the copper was performed for all dopants.
Such a concentration should be high enough to give noticeable effects since it corresponds to a concentration of $10^{18}\un{cm^{-3}}$ while the typical concentration of electrically active defects in undoped Cu$_2$O is always of the order of $10^{15}\un{cm^{-3}}$.
Iron was tried also at higher concentrations.

\section{Experimental results and discussion}

X-ray diffraction measurements do not evidence any peak except
those due to the Cu$_2$O phase. We want also to point out that even if we are not able to measure the impurity distribution inside the samples at these low concentrations, in a highly Ag doped test sample, SEM microanalysis measurements show that the silver is uniformly distributed with a concentration in good agreement with the evaporated metal amount.

From the morphological point of view, the samples show a slight reduction of the grain size with respect to the undoped samples obtained by the same oxidation procedure.
In the undoped substrates the grain size is of the order of $1\un{mm}$ or more and it is obtained after the one hour long step at $1125\un{\C}$.
The doped substrates have instead a grain size of about $100-300\un{\micro m}$, very similar to that of the undoped sample before the grain growth step.
This is likely due to the presence of the impurities which can segregate at the grain boundaries, blocking the grain boundaries diffusion and therefore the grain growth.
However we noticed that in the samples doped with a greater concentration of iron, about $100\un{ppm}$, the grain size increased up to several millimeters. A further increase of the iron concentration decrease the mean grain size. Presently we have no clear explanation of this phenomenon.

\begin{figure}
\centering
\includegraphics[width=0.49\columnwidth]{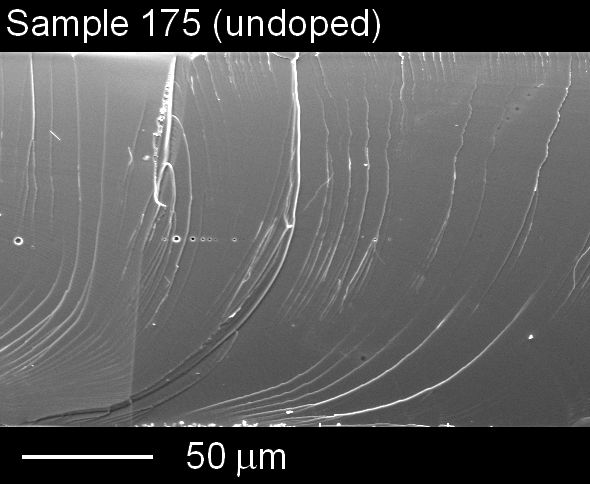}\hfill
\includegraphics[width=0.49\columnwidth]{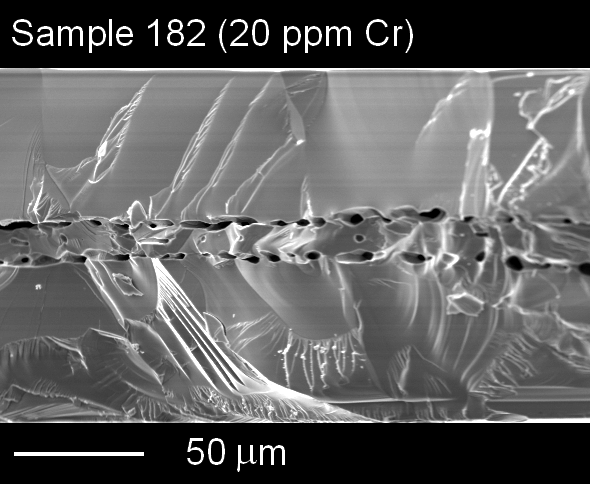}
\caption{SEM images of the cross section of two Cu$_2$O samples: undoped on the left and doped with $20\un{ppm}$ of chromium on the right.}
\label{fig:voids}
\end{figure}

Another morphological consequence of the presence of impurities is an increase of the voids in the middle of the section of the doped Cu$_2$O samples with respect to the undoped ones.
This effect is particularly noticeable in the sample doped with $20\un{ppm}$ of chromium, where the number of voids has increased so much that the two Cu$_2$O scales are almost separated by a central highly defective Cu$_2$O layer.
This is shown in Fig.~\ref{fig:voids} where a cross section of the chromium doped sample together with an undoped one are reported. The same effect is observed in the phosphorus doped sample.
This phenomenon is likely due to the presence of impurities, which can reduce the plasticity of the copper and of the cuprous oxide~\cite{Zhu_2004_B}.

On the other hand the electrical properties of the doped samples are not greatly influenced by the presence of the impurities (see Tab.~\ref{tab:results}). We have measured the resistivity and the mobility in the Van der Pauw configuration by a Bio-Rad HL5900 Hall profiler. 
All our trials do not show any decrease of the resistivity. Chromium, sulfur and phosphorus lead instead to a small increase of the resistivity.
This could be due to the bad morphology of these three sample or to a real change of the compensation ratio.
The mobility is not greatly influenced by the presence of the impurities. Indeed an impurity concentration lower than $10^{18}\un{cm^{-3}}$ is probably unable to decrease a mobility which is already below $100\un{cm^{2}/(V\,s)}$.

For a better characterization, we have fabricated a photoconductor and a solar cell for each substrate.
All these devices have an active area of $0.5\un{cm^2}$ and a back ohmic
contact made by evaporated gold.
Photoconductors were fabricated evaporating a $7\un{nm}$ thick
semitransparent front gold contact covered by an anti-reflection layer
of ZnS ($32\un{nm}$) and by a gold grid.
The solar cells have the same structure with the semitransparent gold substituted by a copper layer of the same thickness.
Just before device preparation, a wet etching procedure ($10\un{s}$ in a $7\un{M}$ HNO$_3$ solution saturated with
NaCl and then $10\un{s}$ in HNO$_3$ (65\%)) has been used to remove a possible thin CuO surface layer.

On the photoconductors we have measured carrier concentration \textit{vs} temperature and persistent photoconductivity decay. From the former measurement we can extract the compensation ratio $N\rb{A}/N\rb{D^+}$ and the energy level of the acceptors, while the latter can give some information on the nature of the donors~\cite{Mittiga_2009}.
During the measurements, the devices were located into a liquid
nitrogen cryostat in vacuum ($P<10^{-5}\un{mbar}$).

Doped samples are very similar to the intrinsic ones:
similar activation energies of the conductivity, similar value of the compensation ratio,
same intensity and decay time constants for the PPC effect.
The absence of doping effects due to the trivalent metals Fe and Cr can be understood considering the numerical simulation reported in the work of Wright~\cite{Wright_2002}.
He found that aluminium and indium in Cu$_2$O easily form neutral complexes with two copper vacancies, becoming electrically inactive.
These complexes have a very large binding energies of the order of $3\un{eV}$.
As for the silver doping the lack of appreciable changes in the conductivity can be easily explained by the fact that silver forms the oxide Ag$_2$O which has exactly the same structure of Cu$_2$O.

The similarity of PPC effect in both doped and undoped samples means that the tested impurities do not affect the mechanism responsible for the PPC at variance with the behavior observed for chlorine doped samples~\cite{Biccari_2010}.

The only clear difference is that doped samples show an enhanced degradation rate of the photoconductivity under illumination with respect to undoped ones where this effect is almost negligible.
This degradation is particularly evident in samples doped with
silicon and phosphorus and it seems to be correlated with a compensation ratio very near to one.

\begin{figure}
\includegraphics[width=\columnwidth]{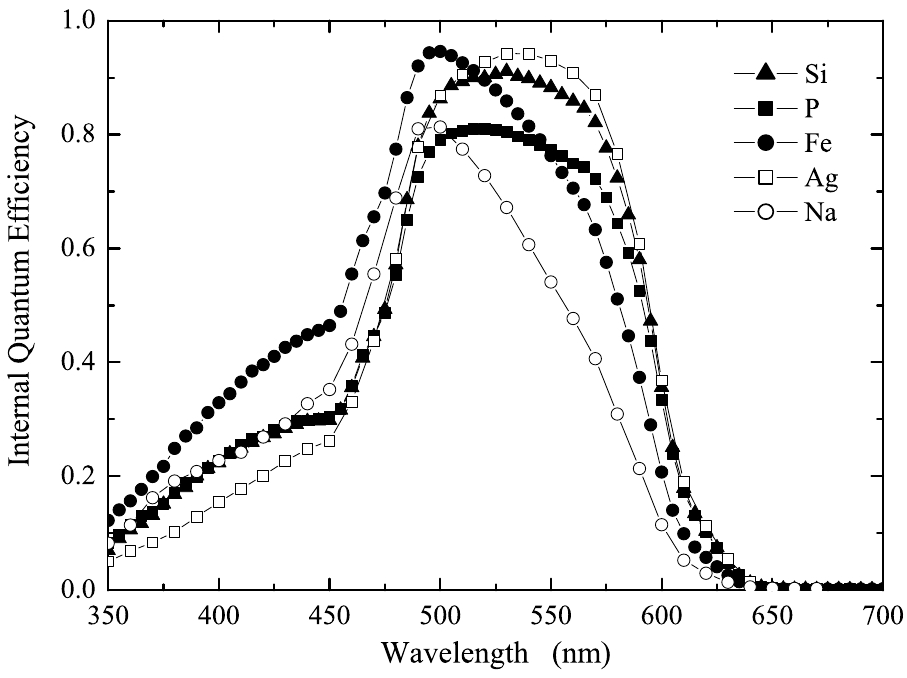}
\caption{Internal quantum efficiencies of solar cells based on the structure ZnS(32\,nm)/Cu(7\,nm)/doped Cu$_2$O/Au(120\,nm).}
\label{fig:IQE}
\end{figure}

From the capacitance--voltage measurement and quantum efficiency on solar cells we can obtain respectively the space charge region width at zero bias, the net charge $N\rb{A}-N\rb{D^+}$ outside the space charge region~\cite{Biccari_2008_proc} and an estimate of the minority carrier diffusion length, $L\rb{n}$~\cite{Olsen_1982}.
All the electron diffusion lengths are about $3\un{\micro m}$ except for sodium doped Cu$_2$O which shows a clear reduction as shown in Fig.~\ref{fig:IQE}.
However this is not a problem because Na is not a typical impurity of the commercial copper.

\begin{table*}
\caption{Mean grain size, hole mobility $\mu\rb{p}$, resistivity (under ambient illumination) $\rho$, conductivity activation energy $E\rb{A}$, compensation ratio $N\rb{A}/N\rb{D^+}$, net charge density $N\rb{A}-N\rb{D^+}$ and width of the space charge region $W$ of several doped Cu$_2$O samples.}
\label{tab:results}
\begin{tabular*}{\textwidth}{lCCCCCCCCl}
\toprule
impurity & \text{level} & \text{grains} & \mu\rb{p} & \rho(300\un{K}) & E\rb{A} & N\rb{A}/N\rb{D^+} & N\rb{A}-N\rb{D^+} & W & notes \\
 & \mathrm{(ppm)} & (\mathrm{\micro m}) & (\mathrm{cm^2/(V\,s)}) & (\mathrm{\ohm\,cm}) & (\mathrm{eV}) & & (10^{14}\,\mathrm{cm^{-3}}) & (\mathrm{\micro m}) & \\
\midrule
---                       & \text{---} & 1000 & 90 & 3000 & 0.33       & 1.2            & 1.5 & 1.8 & \\
Ag                      & 20           &  150   & 83 & 3900 & 0.38       & 2.13          & 1.3 & 2.0 & \\
Cr                       & 20           & 150   & 90 & 5430 & 0.34        & 1.10         & 6.0 & 1.1 &
\small large and dispersed voids\\
Fe                       & 20           & 200   & 92 & 3000 & 0.33        & 1.15         & 2.2 & 1.5 & \\
Fe                       & 100         & 4000 & 99 & 3300 & \text{---} & \text{---} & 2.0 & 1.5 & \\
Fe                       & 1000        & 150  & 94 & 2940 & \text{---} & \text{---} & 1.8 & 2.0 & \\
Na (NaOH)      & 20            & 150   & 85 & 2500 & \text{---} & \text{---} & 3.5 & 0.8 & \small wavy surface \\
P (P$_2$O$_5$) & 20         & 300   & 82 & 6870 & 0.27         & 1.01         & 2.4 & 1.7 & \small large voids \\
S (CuS)            & 20            &  150  & 88 & 4000 & \text{---} & \text{---} & 6.0 & 1.2 & \small inhomogeneous S distribution \\
Si (SiO)            & 20            & 150   & 93 & 3480 & 0.32        & 1.06          & 0.9 & 2.6 & \small dispersed voids \\
\bottomrule
\end{tabular*}
\end{table*}

In conclusion the impurities tested in this work have not shown considerable effects on Cu$_2$O properties.
This confirms all the previous works which found that doping is not able to change the Cu$_2$O conductivity more than one order of magnitude. Probably for photovoltaic applications it will be necessary to use undoped thin film as in CIS case.

\section*{Acknowledgments}
C. Malerba gratefully acknowledges a research grant within the project ``Modified copper oxide for high efficiency photovoltaic cells'' funded by Fondazione Cassa di Risparmio di Trento e Rovereto.
The authors would like to thank Enrico Salza for his skillful technical support and the SEM images.


\end{document}